\documentclass[aps,prx,twocolumn,twoside,superscriptaddress,longbibliography,nofootinbib]{revtex4-2}

\usepackage{amsmath,amssymb,amsfonts}
\usepackage{graphicx}

\usepackage{xstring}

\usepackage{fancyhdr}

\usepackage[colorlinks=true,linkcolor=blue,citecolor=blue,urlcolor=blue,breaklinks=true]{hyperref}
\usepackage{xurl}

\usepackage{algorithm2e}
\usepackage{multirow}
\usepackage{makecell}
\usepackage{longtable}
\usepackage{booktabs}
\usepackage{threeparttable}

\renewcommand{\part}[1]{\left(#1\right)}

\makeatletter
\renewcommand{\@hangfrom@appendix}[3]{%
 #1%
 \@if@empty{#2}{%
  \MakeTextUppercase{#3}%
 }{%
  \MakeTextUppercase{#2}\@if@empty{#3}{}{:\ \MakeTextUppercase{#3}}%
 }%
}%
\makeatother

\usepackage{orcidlink}
\usepackage{marvosym}
\newcommand{\HeaderAuthor}{GRANDA ARANGO \textit{et al.}}
\newcommand{\HeaderTitleFull}{GENERALIZED SVETLINCHNY INEQUALITIES AS A... }
\newcommand{\HeaderTitleMaxChars}{65}

\newcommand{\HeaderTitle}{%
  \StrLen{\HeaderTitleFull}[\HeaderTitleLen]%
  \ifnum\HeaderTitleLen>\HeaderTitleMaxChars
    \StrLeft{\HeaderTitleFull}{60}\ldots
  \else
    \HeaderTitleFull
  \fi
}

\fancypagestyle{prxquantum}{%
  \fancyhf{}%
  \fancyhead[LE]{\small\slshape \HeaderAuthor}
  \fancyhead[LO]{\small\slshape \HeaderTitle}
  \fancyfoot[C]{\thepage}%
}

\begin{document}

\title{Generalized Svetlichny Inequalities as a Cross-Platform Benchmark for Quantum Processors}


\author{Andr\'es Camilo Granda Arango\orcidlink{0000-0002-5255-2507}}
\email{Contact author: a.grandaarango@gmail.com}
\affiliation{\href{https://ror.org/003109y17}{Universit\`a degli Studi di Cagliari}, Via Is Mirrionis 1, Cagliari, 09123, Italy}

\author{Giuseppe Sergioli\orcidlink{0000-0002-3650-5858}}
\email{Contact author: giuseppe.sergioli@gmail.com}
\affiliation{\href{https://ror.org/003109y17}{Universit\`a degli Studi di Cagliari}, Via Is Mirrionis 1, Cagliari, 09123, Italy}

\author{Luigi Iapichino\orcidlink{0000-0003-3938-8973}}
\affiliation{\href{https://ror.org/05558nw16}{Leibniz Supercomputing Centre of the Bavarian Academy of Sciences and Humanities (LRZ)}, Boltzmannstr. 1, 85748 Garching b. M\"unchen, Germany}

\author{Tobias Valentin Bauer\orcidlink{0009-0004-8511-0563}}
\affiliation{\href{https://ror.org/05558nw16}{Leibniz Supercomputing Centre of the Bavarian Academy of Sciences and Humanities (LRZ)}, Boltzmannstr. 1, 85748 Garching b. M\"unchen, Germany}

\author{Carlo Cuccu\orcidlink{0000-0002-4519-1854}}
\affiliation{\href{https://ror.org/003109y17}{Universit\`a degli Studi di Cagliari}, Via Is Mirrionis 1, Cagliari, 09123, Italy}

\author{Roberto Giuntini\orcidlink{0000-0003-2078-0297}}
\affiliation{\href{https://ror.org/003109y17}{Universit\`a degli Studi di Cagliari}, Via Is Mirrionis 1, Cagliari, 09123, Italy}
\affiliation{\href{https://ror.org/02kkvpp62}{Institute for Advanced Study (IAS), Technische Universit\"at M\"unchen}, Lichtenbergstra{\ss}e 2a, 85748 M\"unchen, Germany}

\begin{abstract}
A cross-platform benchmark should be device-independent yet sensitive to imperfections accumulated during execution. Building on prior work, we formulate and validate a protocol through demonstrations based on generalized Svetlichny inequality violations (a simulation, rather than a loophole-free certification, of multipartite nonlocality) for larger subsystem sizes. We apply it unchanged to three processors, the trapped-ion \texttt{Marmot} and \texttt{IBEX~Q1} (both manufactured by AQT) and the superconducting \texttt{Euro-Q-Exa System~1} (IQM), using identical GHZ-state circuits with fixed theoretical measurement settings from $N=2$ to $N=10$ qubits. Both trapped-ion processors sustain violations up to ten qubits, while the superconducting processor loses them beyond six qubits under unmitigated (raw) conditions; readout-error mitigation recovers violations up to eight qubits, with a substantial fraction of the degradation attributable to measurement imperfections. These processors provide a heterogeneous testbed for validating the protocol's cross-platform applicability and discriminating power, demonstrating its practicality as a common framework for assessing collective quantum performance.
\end{abstract}


\maketitle

\thispagestyle{empty}
\pagestyle{prxquantum}

\section{Introduction}
\label{sec:introduction}

Quantum computing has experienced remarkable progress during the last decade, leading to the emergence of increasingly sophisticated noisy intermediate-scale quantum (NISQ) processors based on a variety of physical platforms, including superconducting circuits~\cite{SuperconductingQubits_review}, trapped ions~\cite{IonTrapQuantumComputing_review}, neutral atoms~\cite{Henriet2020}, and photonic architectures~\cite{Madsen_Xanadu_2022}, alongside growing efforts toward fault-tolerant quantum computing (FTQC). As these technologies continue to scale, the problem of reliably characterizing and comparing quantum processors has become one of the central challenges in the roadmap toward practical quantum advantage. In this context, benchmarking protocols play a fundamental role, not only for evaluating hardware quality, but also for diagnosing limitations associated with noise, connectivity, compilation overhead, calibration drift, and measurement imperfections.

Most currently adopted benchmarking methodologies are based on local or quasi-local quantities, such as single- and two-qubit gate fidelities, randomized benchmarking~\cite{Magesan2011}, cross-entropy benchmarking~\cite{Arute_2019_Sycamore}, quantum volume~\cite{Cross2019}, or algorithmic success probabilities. Although these quantities provide valuable information about isolated components of a quantum processor, they do not necessarily characterize the collective quantum behavior of the system as a whole. In particular, high local fidelities do not guarantee that a processor is capable of generating, preserving, and manipulating large-scale multipartite quantum correlations under realistic execution conditions. This distinction becomes increasingly relevant as quantum hardware grows in size and complexity.
A cross-platform benchmark of such collective behavior should therefore satisfy two complementary requirements: it should be defined at the logical level independently of the target hardware, while remaining sufficiently sensitive to the device-dependent imperfections introduced during compilation and physical execution. This separation between a common benchmark specification and its hardware-specific implementation provides a natural basis for comparing collective performance across heterogeneous quantum processors.

From a physical perspective, multipartite quantum correlations constitute emergent collective properties that cannot be reduced to the independent performance of individual qubits or isolated gate operations. Multipartite nonlocality, as characterized through the generalized Svetlichny inequalities, represents one of the strongest manifestations of this collective quantum behavior. Violations of this bound witness multipartite correlations incompatible with a hybrid local-realism (HLR) description. More refined operational definitions of genuine multipartite nonlocality have since been proposed~\cite{Gallego2012,Bancal2013}; we return to their precise relation to the Svetlichny criterion in Background (Sec.~\ref{sec:GMNL}). In this work, Svetlichny violations are adopted as an experimentally demanding and highly sensitive benchmark of collective multipartite correlations in quantum processors.

This intrinsic sensitivity makes Svetlichny violations particularly attractive as a benchmarking resource. However, they also introduce substantial experimental and computational challenges. The complexity of multipartite Bell operators grows exponentially with the number of qubits, while the corresponding violations become increasingly fragile under decoherence, imperfect entangling gates, crosstalk, measurement noise, and transpilation overhead. As a result, preserving Svetlichny violations at increasing subsystem sizes becomes a demanding task for present-day NISQ architectures. Precisely because of this fragility, multipartite nonlocality tests offer a highly sensitive probe of the global performance of quantum processors.

Because the qubits of a quantum processor are co-located and controlled within a single device, measurements on different qubits cannot be arranged to be spacelike separated. Our tests cannot therefore provide a loophole-free certification of genuine multipartite nonlocality (GMNL) in the Bell-test sense. We use the expression \emph{simulating nonlocality} to describe what our protocol does instead, in the sense made precise in Sec.~\ref{sec:GMNL}.

While multipartite Bell inequalities have been explored in a variety of experimental and computational settings~\cite{Granda_2025,Bonsel2024,Kam2024,Lanyon2014,Baumer2021}, systematic studies of their scalability across different quantum computing architectures under a common benchmarking protocol remain relatively limited. Recent work has demonstrated n-partite Bell nonlocality on real quantum processors for increasing system sizes~\cite{Bonsel2024}, and characterized how GHZ-state fidelity and multipartite entanglement degrade with subsystem size on superconducting hardware~\cite{Kam2024,Wang2025}. Unlike these proposals, which adapt the resource state to device connectivity and
estimate the Bell operator from a randomly sampled subset of measurement
contexts~\cite{Bonsel2024,Wang2025}, we reconstruct every individual correlator
of the generalized Svetlichny inequality directly on real hardware, using an
identical, connectivity-independent logical circuit at every subsystem
size.\footnote{Bönsel, Gühne, and Cabello~\cite{Bonsel2024} extrapolate
multipartite nonlocality to $n=24$ qubits on a simulated IBM Eagle processor
using measured device noise rates, rather than direct hardware execution;
moreover, their Bell operator is of Mermin type, which excludes only fully
local hidden-variable models, whereas the Svetlichny criterion adopted here
excludes the strictly larger class of hybrid local-realistic
models~\cite{Granda_2025}, so the two protocols certify distinct,
non-equivalent notions of multipartite nonlocality. Ref.~\cite{Wang2025}
similarly relies on a reduced set of measurement settings that simplify the
Svetlichny operator.}

In this work, we carry out Svetlichny-inequality-based benchmarking across three quantum processors spanning two fundamentally different platforms: the trapped-ion processors \texttt{Marmot}~\cite{AQT_Marmot} and \texttt{IBEX~Q1}~\cite{AQT_IBEX}, both developed by Alpine Quantum Technologies (AQT), and the superconducting processor \texttt{Euro-Q-Exa System~1} (EQE1)~\cite{LRZ_EuroQExa} developed by IQM. Using identical logical GHZ-state circuits, automatically transpiled into the native gate sets of each platform, we study the preservation of multipartite nonlocal correlations from 2 up to 10 qubits, which at the largest size requires the evaluation, via demonstrations, of the complete set of $2^{10}=1,024$ correlators. The study extends the protocol beyond the few-qubit regime and complements previous work on multipartite Bell correlations and on the growth of GHZ-state fidelity and entanglement with qubit number. Rather than optimizing the measurement settings independently for each device, we perform all demonstrations using the theoretical measurement angles prescribed by the original Svetlichny formalism; this keeps the benchmarking protocol consistent and independent of the underlying architecture. These three processors thus constitute a cross-platform testbed for demonstrating that the benchmarking protocol can be consistently implemented, scaled, and compared across distinct hardware architectures.

In addition to the unmitigated (raw) demonstration executions, we investigate the role of measurement noise through the implementation of local and correlated readout error mitigation strategies. This enables us to assess the impact of readout errors on the observed Svetlichny violations and to determine the extent to which measurement imperfections contribute to the degradation of multipartite nonlocal correlations. Furthermore, by comparing three processors across two architectures with fundamentally different connectivity, native gate implementations, and transpilation requirements, our results provide insight into how hardware design influences the scalability of collective quantum correlations.

\section{BACKGROUND}
\label{sec:methods}

\subsection*{Genuine multipartite nonlocality}
\label{sec:GMNL}

The characterization of multipartite nonlocal correlations in large quantum systems can be formulated through the generalized Svetlichny inequalities~\cite{Svetlichny_PRD, Seevinck_Svetlichny_PRL}, which are designed to exclude hybrid local-realism models in multipartite scenarios. For an $N$-qubit system, the generalized Svetlichny operator is~\cite{Seevinck_Svetlichny_PRL}:

\begin{equation}
\label{e:Svelichny_N_Qubits_operator}
    S_{N}^{\pm}=\sum_{I}\nu_{t(I)}^{\pm}\,A^{i_{1}}_{1}\cdots A^{i_{N}}_{N},
\end{equation}

\noindent where $I=(i_1,i_2,\dots,i_N)$ labels the measurement configuration, $t(I)$ denotes the number of primed observables in the correlator, and the sign coefficients are given by $\nu_t^{\pm}=(-1)^{t(t\pm 1)/2}$. Under the hybrid local-realistic assumption, all $N$-partite correlations satisfy the Svetlichny bound

\begin{equation}
\label{e:SVL_N_Qubits}
    |S_{N}^{\pm}|\leq 2^{N-1}.
\end{equation}

Violations of this bound witness multipartite nonlocal correlations incompatible with Svetlichny's hybrid local-realism framework. Alternative operational definitions of genuine multipartite nonlocality have since been proposed, based on hidden-variable models whose bipartite terms are constrained to be non-signaling or time-ordered one-way signaling~\cite{Gallego2012,Bancal2013}. For the tripartite case, both works show that the resulting local sets are strictly smaller than the Svetlichny-local set, so that a Svetlichny violation is sufficient, though not necessary, for genuine multipartite nonlocality under these refined definitions; Bancal \textit{et al.}~\cite{Bancal2013} give an explicit example, a noisy $W$ state, that is genuinely multipartite nonlocal under their definition without violating the Svetlichny inequality. We adopt the generalized, $N$-partite Svetlichny violation here as a conservative and experimentally demanding benchmark of strong multipartite nonclassical correlations.

As mentioned in the Introduction, the correlations associated to the Svetlichny inequality considered in this work are evaluated on quantum processors where qubits are confined in a reduced space. No space-like separation is granted for measurements involving different qubits, and therefore, avoiding the locality loophole on quantum processors remains a challenging issue~\cite{Aspect1982,Hensen_2015,Significant-Loophole-Free-2015,Rosenfeld2017}. For this reason, the protocol based on the Svetlichny inequality cannot be used to certify nonlocality in a loophole-free sense. This limitation should be understood as inherent to the platform rather than as a gap expected to close with incremental hardware improvements: unlike detection efficiency or gate fidelity, which can in principle be improved through better fabrication and control, the absence of spacelike separation follows directly from operating all qubits within a single physical package, and is therefore unlikely to be resolved by near-term advances in processor design. As discussed in Ref.~\cite{Granda_2025} (Sec.~VI.B), the violation of nonlocality inequalities on a QPU can be better understood as a contextuality phenomenon, since nonlocality inequalities are a particular case of noncontextuality inequalities~\cite{Howard2014}. Following Ref.~\cite{Granda_2025}, we speak of \textit{simulating nonlocality} to describe what the proposed test does. This means to reproduce the correlations associated to all the terms involved in the Svetlichny inequality on a quantum processor. As we show in this work, it turns out that simulating those correlations turns out to be a hard task, that can be used to assess the collective behavior of different subsets of qubits. The test is so stringent, that it allows to detect differences in performance between quantum processors even when a small number of qubits is involved.

As the subsystem size increases, maintaining a Svetlichny violation becomes progressively more demanding. Decoherence, crosstalk, gate imperfections, and readout noise all contribute to suppressing $\langle S_{N}^{\pm} \rangle$ below the violation threshold, while the number of correlators required to evaluate the operator grows exponentially with $N$. This sensitivity to hardware imperfections is precisely what makes Svetlichny violations a stringent and informative probe of the collective quantum performance of a processor at scale.

Because the classical bound in Eq.~\ref{e:SVL_N_Qubits} itself grows exponentially with $N$, comparing raw values of $S_N$ across different subsystem sizes is not straightforward. To enable a direct comparison across all subsystem sizes and architectures, we define the normalized Svetlichny violation

\begin{equation}
\label{e:V_N_normalized}
\mathcal{V}_N = \frac{S_N}{2^{N-1}},
\end{equation}

\noindent where $\mathcal{V}_N > 1$ indicates a genuine Svetlichny violation and $\mathcal{V}_N = 1$ corresponds to the HLR threshold. This normalization removes the exponential scaling of the classical bound, placing all processors and subsystem sizes on a common scale.


\section{Results}
\label{sec:results}

\subsection{Noise model simulations}

\begin{figure*}[]
\centering
\includegraphics[width=0.9\linewidth]{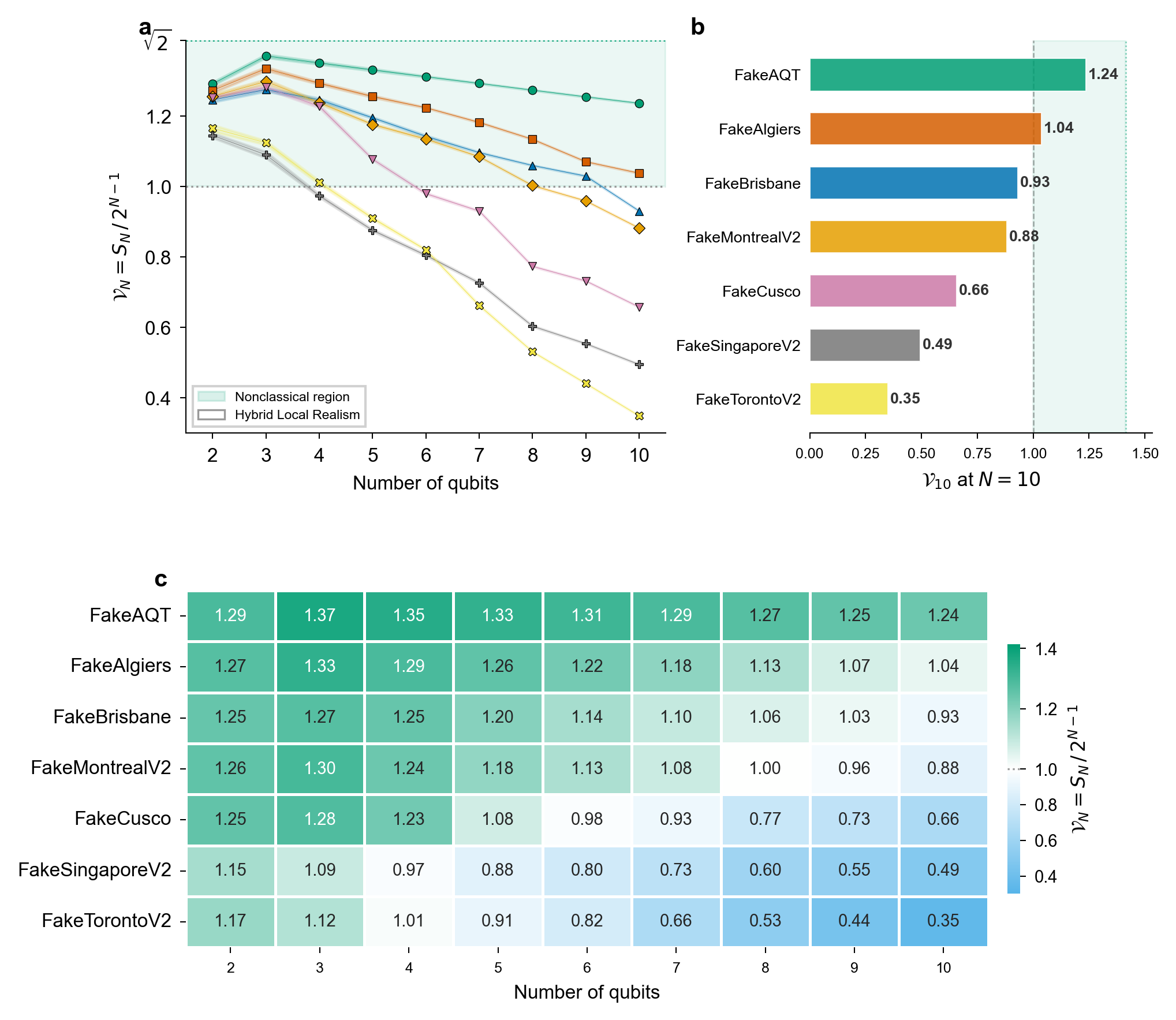}
\caption{Normalized Svetlichny violation $\mathcal{V}_N$ under calibrated hardware noise models for the AQT trapped-ion platform (\texttt{FakeAQT}) and six superconducting noise models available through the Qiskit \texttt{FakeProvider} framework (\texttt{FakeAlgiers}, \texttt{FakeBrisbane}, \texttt{FakeMontrealV2}, \texttt{FakeCusco}, \texttt{FakeSingaporeV2}, \texttt{FakeTorontoV2}). (a) $\mathcal{V}_N$ as a function of the number of qubits, from $N=2$ to $N=10$. Shaded bands around each curve indicate 95\% confidence intervals (based on Student's $t$-distribution) from 20 repetitions per data point, not visible where smaller than the line width. The dotted line at $\mathcal{V}_N=1$ marks the boundary between the white HLR region and the green nonclassical region. (b) $\mathcal{V}_N$ at $N=10$ for each simulated backend, ranked in descending order. (c) Full numerical values of $\mathcal{V}_N$ for each backend.}
\label{fig:Simulation}
\end{figure*}

Before analyzing the demonstration data, we evaluate how generalized Svetlichny violations behave under realistic hardware noise models as the subsystem size increases to establish a reference framework for interpretation. In this section, we report the normalized Svetlichny violation $\mathcal{V}_N = S_N/2^{N-1}$ (Eq.~\ref{e:V_N_normalized}, Background), which removes the exponential growth of the classical bound and enables a direct comparison of performance across subsystem sizes. Fig.~\ref{fig:Simulation} compares the ideal theoretical prediction with simulations performed using calibrated backend models corresponding to the AQT trapped-ion platform and several superconducting noise models available through the Qiskit \texttt{FakeProvider} framework~\cite{Qiskit}. For consistency with the demonstration measurements, all simulated data points were obtained using 2,000 shots (see Appendix~\ref{sec:hardware}).

All simulated backends exhibit a progressive decrease in the Svetlichny violation with subsystem size. The AQT model (\texttt{FakeAQT}) tracks the theoretical prediction most closely, remaining above the HLR threshold across the entire explored range. Of the seven simulated backends, only \texttt{FakeAQT} and \texttt{FakeAlgiers} remain within the nonclassical regime up to $N=10$, while the five remaining superconducting backends (\texttt{FakeBrisbane}, \texttt{FakeMontrealV2}, \texttt{FakeCusco}, \texttt{FakeSingaporeV2}, \texttt{FakeTorontoV2}) cross into the HLR regime before reaching $N=10$, reflecting the diversity of noise levels and connectivity constraints across device generations. This dispersion qualitatively anticipates the behavior observed experimentally in Fig.~\ref{fig:Scaling_10Qubits}, suggesting that sustaining the violation up to $N=10$ is achievable on real hardware under noise levels comparable to those of the least noisy models, as confirmed by the demonstration data in the following section.

 \subsection{Implementation in quantum processors}
\begin{figure}[]
\centering
\includegraphics[width=\linewidth]{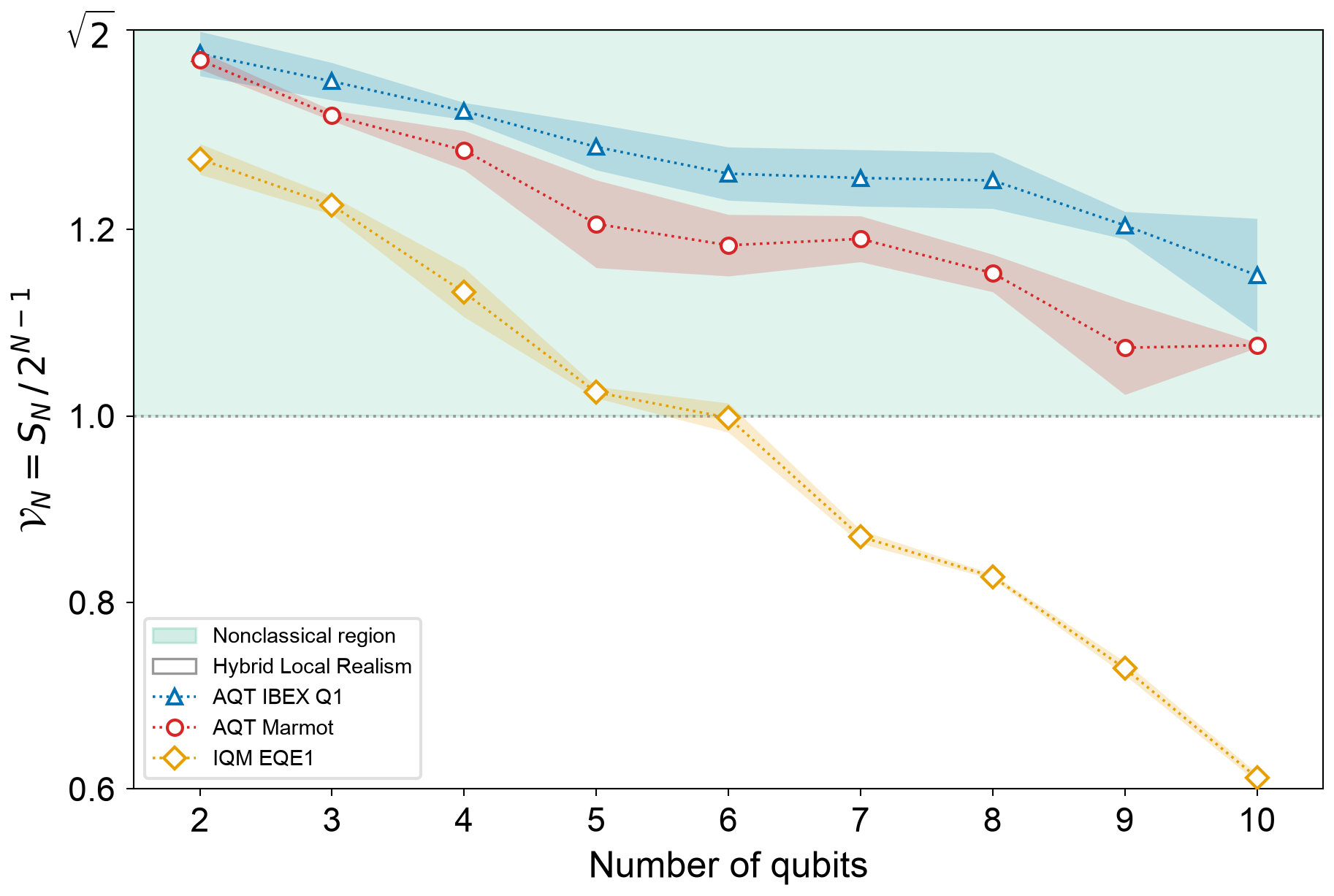}
\caption{Normalized Svetlichny violation $\mathcal{V}_N$ from $N=2$ to $N=10$ qubits. Blue triangles correspond to \texttt{AQT~IBEX~Q1}, red circles to \texttt{AQT~Marmot}, and orange diamonds to \texttt{IQM~EQE1}. Shaded bands around each curve indicate 95\% confidence intervals. The dotted line at $\mathcal{V}_N=1$ marks the boundary between the white HLR region and the green nonclassical region.}

\label{fig:Scaling_10Qubits}
\end{figure}

We now turn to the demonstration execution of this protocol on real hardware, following the quantum processors and GHZ-circuit implementation described in Appendix~\ref{sec:hardware} and~\ref{sec:ghz}. The results from the demonstrations, shown in Fig.~\ref{fig:Scaling_10Qubits}, reveal a pronounced divergence in performance between the trapped-ion and superconducting architectures.

Both trapped-ion processors, \texttt{AQT~Marmot} and \texttt{AQT~IBEX~Q1}, sustain statistically significant Svetlichny violations across the complete explored range up to $N=10$ qubits, remaining within the nonclassical region throughout. \texttt{AQT~IBEX~Q1} consistently attains values closer to the ideal theoretical maximum ($\sqrt{2}$) than \texttt{Marmot} across most of the explored range, although the two processors' confidence intervals converge toward $N=9$--$10$. Marmot's values (from the demonstrations) track the AQT simulator predictions closely only at the smallest subsystem sizes ($N=2$--$3$); beyond this range a persistent gap in $\mathcal{V}_N$ emerges and remains roughly constant in magnitude up to $N=10$, consistent with the increasing sensitivity of the Svetlichny operator to hardware imperfections at larger subsystem sizes. 

The superconducting processor \texttt{IQM~EQE1} exhibits a qualitatively different behavior: while all three processors show comparable performance at the smallest subsystem sizes ($N=2$--$3$), \texttt{IQM~EQE1} crosses the HLR bound at $N=6$ and its violations continue degrading well below the classical bound for $N\geq7$ under raw execution conditions. Since all three processors are evaluated using identical logical circuits and the same theoretical measurement prescription, the pronounced gap between the trapped-ion and superconducting platforms reflects the combined influence of hardware coherence, connectivity, native gate implementation, and transpilation overhead intrinsic to each architecture~\footnote{The corresponding compilation overhead for a representative measurement circuit is quantified in Table~\ref{tab:circuit_complexity}, which compares the logical circuit depth and gate counts with the transpiled implementations for the IQM and AQT native gate sets.}, rather than any optimization advantage of one platform over the other. 
These results demonstrate that simulating genuine multipartite nonlocality, in the sense discussed in Sec.~\ref{sec:introduction}, remains a highly informative benchmark even under real, noisy execution conditions on current NISQ hardware.

\begin{table*}[]
\centering
\footnotesize
\caption{Compilation overhead for a representative measurement circuit used to evaluate the generalized Svetlichny inequality as a function of subsystem size $N$. For each subsystem size, the logical circuit (submitted) is compared with its transpiled implementation for the IQM and AQT native gate sets. The logical circuit is architecture-independent (identical for both platforms) and is therefore reported only once.}
\label{tab:circuit_complexity}
\resizebox{\textwidth}{!}{%
\begin{tabular}{c ccc ccc ccc ccc}
\toprule
 & \multicolumn{3}{c}{Depth} & \multicolumn{3}{c}{1-qubit gates} & \multicolumn{3}{c}{2-qubit gates} & \multicolumn{3}{c}{Gate types} \\
\cmidrule(lr){2-4} \cmidrule(lr){5-7} \cmidrule(lr){8-10} \cmidrule(lr){11-13}
$N$ & Submitted & IQM & AQT & Submitted & IQM & AQT & Submitted & IQM & AQT & Submitted & IQM & AQT \\
\midrule
2  & 5  & 4  & 5  & 5  & 4  & 4  & 1 & 1 & 1 & cx:1, h:1, ry:2, rz:2  & cz:1, r:4  & r:3, rxx:1, rz:1 \\
3  & 6  & 6  & 8  & 7  & 7  & 7  & 2 & 2 & 2 & cx:2, h:1, ry:3, rz:3  & cz:2, r:7  & r:5, rxx:2, rz:2 \\
4  & 7  & 8  & 11 & 9  & 10 & 10 & 3 & 3 & 3 & cx:3, h:1, ry:4, rz:4  & cz:3, r:10 & r:7, rxx:3, rz:3 \\
5  & 8  & 10 & 14 & 11 & 13 & 13 & 4 & 4 & 4 & cx:4, h:1, ry:5, rz:5  & cz:4, r:13 & r:9, rxx:4, rz:4 \\
6  & 9  & 12 & 17 & 13 & 16 & 16 & 5 & 5 & 5 & cx:5, h:1, ry:6, rz:6  & cz:5, r:16 & r:11, rxx:5, rz:5 \\
7  & 10 & 14 & 20 & 15 & 19 & 19 & 6 & 6 & 6 & cx:6, h:1, ry:7, rz:7  & cz:6, r:19 & r:13, rxx:6, rz:6 \\
8  & 11 & 16 & 23 & 17 & 22 & 22 & 7 & 7 & 7 & cx:7, h:1, ry:8, rz:8  & cz:7, r:22 & r:15, rxx:7, rz:7 \\
9  & 12 & 18 & 26 & 19 & 25 & 25 & 8 & 8 & 8 & cx:8, h:1, ry:9, rz:9  & cz:8, r:25 & r:17, rxx:8, rz:8 \\
10 & 13 & 20 & 29 & 21 & 28 & 28 & 9 & 9 & 9 & cx:9, h:1, ry:10, rz:10 & cz:9, r:28 & r:19, rxx:9, rz:9 \\
\bottomrule
\end{tabular}%
}

\vspace{2mm}
\footnotesize Gate name abbreviations: \texttt{h} = Hadamard, \texttt{cx} = CNOT, \texttt{ry}/\texttt{rz} = single-qubit rotations, \texttt{cz} = controlled-Z (IQM native), \texttt{r} = generic single-qubit rotation, \texttt{rxx} = M{\o}lmer--S{\o}rensen-type two-qubit rotation (AQT native).
\end{table*}


\subsection{Error mitigation}

\begin{figure*}[t]
\centering
\includegraphics[width=\linewidth]{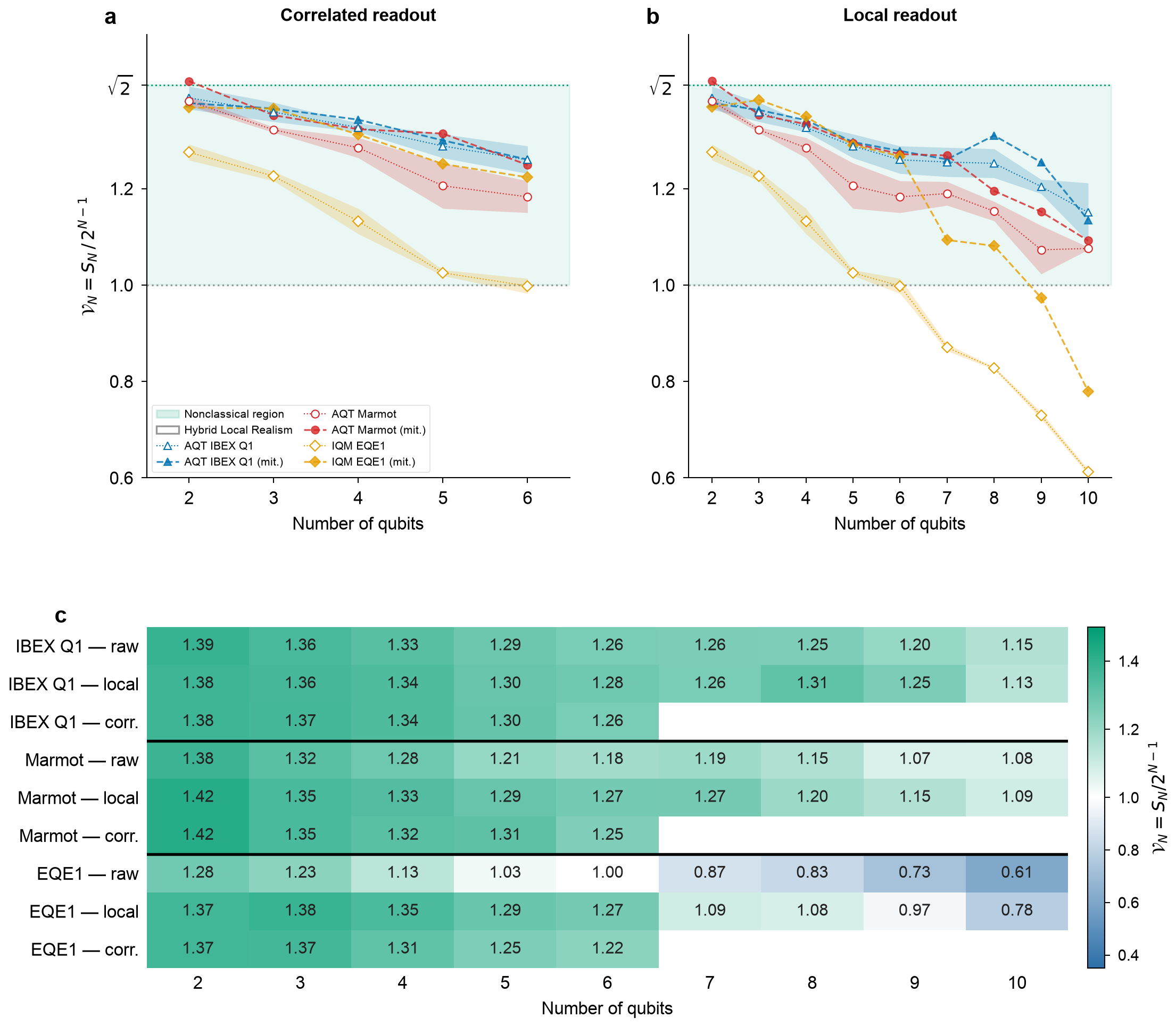}
\caption{Readout error mitigation. Normalized Svetlichny violation $\mathcal{V}_N$ for \texttt{AQT~IBEX~Q1} (blue triangles), \texttt{AQT~Marmot} (red circles), and \texttt{IQM~EQE1} (orange diamonds) under unmitigated (open markers) and mitigated (solid markers) conditions. (a) Correlated readout mitigation up to $N=6$. The dotted green line at $\mathcal{V}_N=\sqrt{2}$ indicates the ideal theoretical prediction for GHZ states; the dotted line at $\mathcal{V}_N=1$ marks the boundary between the white HLR region and the green nonclassical region. Shaded bands indicate 95\% confidence intervals. (b) Local readout mitigation up to $N=10$. (c) Corresponding numerical values for all three processors and mitigation strategies.}
\label{fig:Normalized}
\end{figure*}

To determine the extent to which these differences originate from measurement imperfections rather than from genuine decoherence of the underlying multipartite state, we applied local and correlated readout error mitigation strategies to the demonstration data: the former assumes independent single-qubit assignment errors and remains tractable up to $N=10$, whereas the latter captures joint readout errors at the cost of an exponentially growing calibration overhead, restricting it to $N\leq6$ (full protocol in Appendix~\ref{sec:mitigation}). To facilitate a direct visual comparison across all subsystem sizes, we use the normalized Svetlichny violation $\mathcal{V}_N$ introduced in  (Eq.~\ref{e:V_N_normalized}, Sec.~\ref{sec:GMNL}), which removes the exponential scaling of the classical bound and places all processors and subsystem sizes on a common scale. The results are shown in Fig.~\ref{fig:Normalized}.

\begin{figure*}[]
\centering
\includegraphics[width=0.8\linewidth]{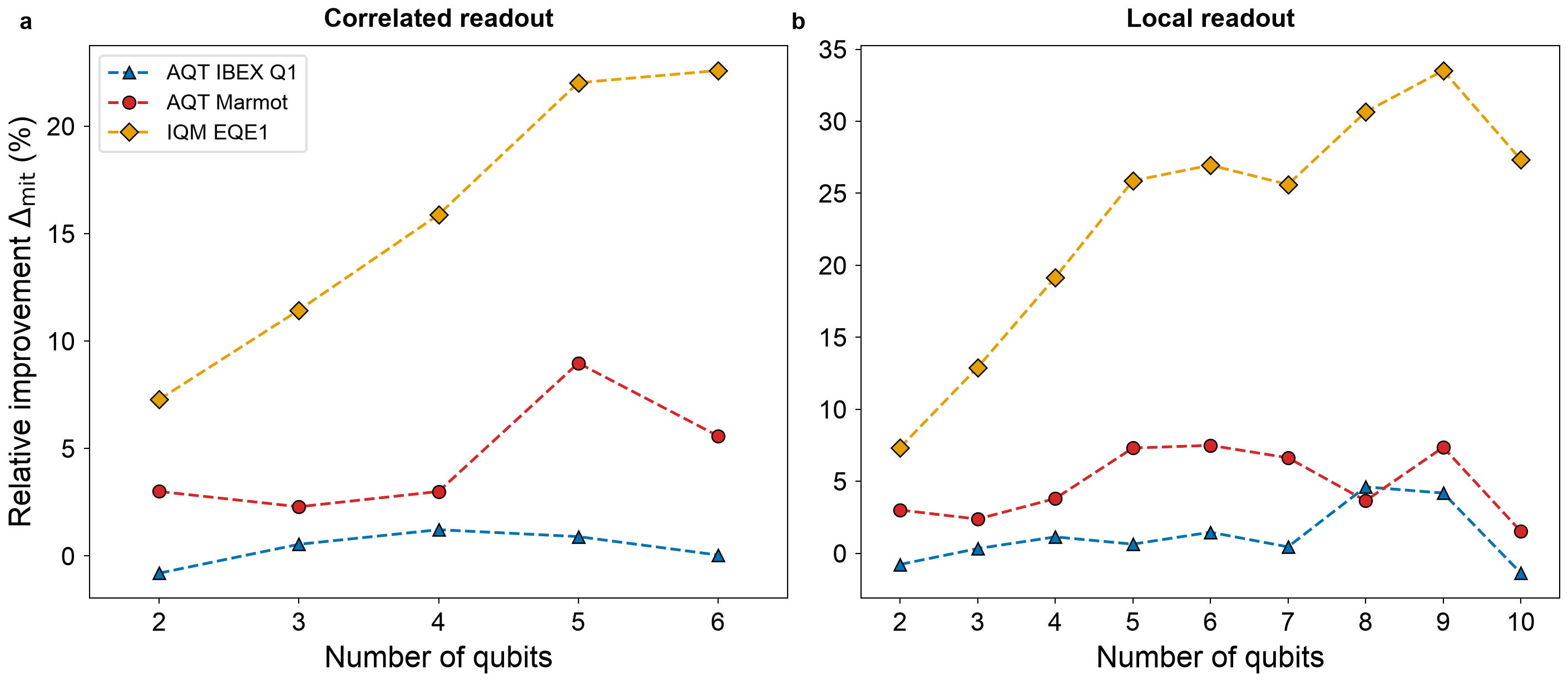}
\caption{Mitigation gain. Relative improvement from readout error mitigation, $\Delta_{\mathrm{mit}}$ (\%), for \texttt{AQT~IBEX~Q1}, \texttt{AQT~Marmot}, and \texttt{IQM~EQE1}. (a) Correlated readout mitigation, up to $N=6$. (b) Local readout mitigation, up to $N=10$.}
\label{fig:MitigationHeatmap}
\end{figure*}

The mitigation analysis reveals a striking asymmetry between the trapped-ion and superconducting architectures. For both \texttt{AQT~Marmot} and \texttt{AQT~IBEX~Q1}, local and correlated mitigation produce only modest corrections throughout the explored range, with relative improvements remaining below approximately 9\% at every subsystem size for both processors (see Fig.~\ref{fig:MitigationHeatmap}). This indicates that readout errors are not the dominant source of degradation on either trapped-ion platform, and that the observed violations are primarily limited by intrinsic decoherence and gate imperfections affecting the multipartite state itself, a behavior consistent across both AQT devices despite their different raw performance. For \texttt{IQM~EQE1}, the behavior is qualitatively different: the relative improvement from mitigation (Eq.~\ref{e:delta_mit}) grows rapidly with subsystem size, reaching approximately 27\% at $N=10$, substantially larger than the corrections observed for either AQT processor. This strong recovery demonstrates that a substantial fraction of the apparent degradation observed under raw execution conditions in the superconducting architecture originates from measurement imperfections rather than from a complete destruction of the underlying multipartite correlations.

These observations allow us to directly compare the collective quantum performance of the two architectures in terms of the largest subsystem size for which statistically significant Svetlichny violations remain  observable. Under raw execution conditions, \texttt{AQT~Marmot} sustains violations throughout the complete explored range up to $N=10$, while \texttt{IQM~EQE1} loses observable nonlocality beyond approximately $N=6$. After applying local readout mitigation, the superconducting platform recovers Svetlichny violations up to approximately $N=8$ qubits, substantially extending its accessible nonclassical region.

The gap between the raw and mitigated accessible ranges in \texttt{IQM~EQE1} provides direct evidence from the demonstrations that measurement imperfections contribute substantially to the observed degradation, whereas their impact is comparatively small for the AQT processors. Importantly, this distinction cannot be inferred from readout performance alone: it emerges from the way measurement errors combine with state preparation, gate errors, and decoherence in the multipartite regime. The results therefore highlight that similar local gate fidelities do not necessarily translate into comparable collective quantum performance, emphasizing the complementary information provided by multipartite nonlocality benchmarking.

\section{Discussion}
\label{sec:discussion}

The results presented in this work demonstrate that generalized Svetlichny inequalities provide a highly sensitive probe of collective quantum performance in current NISQ processors, capturing system-level properties that are not directly resolved by conventional local benchmarking metrics. Building on previous theoretical and experimental developments~\cite{Granda_2025}, this work establishes generalized Svetlichny benchmarking as a scalable, cross-platform protocol whose logical definition is independent of the underlying hardware architecture. The three processors examined here provide a heterogeneous demonstration testbed for demonstrating that the same benchmarking procedure can be consistently implemented and scaled across distinct quantum computing architectures, while retaining its ability to discriminate differences in collective performance. Several observations emerging from this comparison deserve closer examination and further illustrate the diagnostic power of the protocol.

The most immediate result is the pronounced platform- and device-dependent fragility of multipartite nonlocality. Both trapped-ion processors, \texttt{AQT~Marmot} and \texttt{AQT~IBEX~Q1}, sustain statistically significant Svetlichny violations across the complete explored range up to $N=10$ qubits, while the superconducting processor \texttt{IQM~EQE1} loses observable nonlocality beyond approximately $N=6$ under raw execution conditions. This difference cannot be attributed to state preparation optimization or measurement angle tuning, since all three processors were evaluated under identical logical circuits and theoretical prescriptions. Instead, is consistent with a combined influence of connectivity, native gate implementation, transpilation, and hardware-specific noise on the preservation of high-order multipartite correlations. Notably, this overhead is not simply a matter of gate count: as shown in Table~\ref{tab:circuit_complexity}, the AQT and IQM transpiled circuits require an identical number of single- and two-qubit gates at every subsystem size, yet the AQT implementation is consistently deeper~\footnote{This asymmetry originates at the compiler level: on AQT, each pair of chained two-qubit \texttt{RXX} gates is separated by a physical \texttt{RZ} correction that cannot be algebraically absorbed into an adjacent single-qubit rotation, whereas the IQM compiler merges the equivalent phase correction into the neighboring gate through virtual-$Z$ tracking, effectively parallelizing operations that remain sequential on AQT and reducing the critical path length.}. Nevertheless, the trapped-ion processors sustain substantially larger normalized Svetlichny violations than IQM \texttt{EQE1}. This observation indicates that the preservation of high-order multipartite correlations depends on the combined effect of the native gate implementation, gate fidelity, circuit scheduling, and connectivity, rather than on gate count or transpiled depth considered in isolation. In particular, all-to-all connectivity avoids the additional routing overhead that would otherwise arise in connectivity-limited architectures, while differences in native gate decomposition and execution may determine the effective impact of the resulting circuit depth~\cite{Linke2017} on the collective correlations.

Reaching $N=10$ is significant not merely as an increase in subsystem size. At this scale, evaluating the complete generalized Svetlichny operator requires $1,024$ distinct multipartite correlators, each reconstructed at this scale in the demonstrations. The persistence of a statistically significant violation on both trapped-ion processors therefore shows that the benchmark remains discriminating in practice in a regime where the measurement cost and the sensitivity to accumulated imperfections have increased substantially. At the same time, the pronounced degradation observed on \texttt{IQM~EQE1} demonstrates that the large-$N$ regime amplifies device-dependent limitations that remain less visible at smaller subsystem sizes.

A central aspect of this work is the distinction between collective and local quantum performance. Standard benchmarking quantities, such as single- and two-qubit gate fidelities, randomized benchmarking, and quantum volume, characterize different aspects of hardware performance, but do not necessarily reflect the processor's capacity to preserve the high-order multipartite correlations required to observe violations of the Svetlichny inequalities. The Svetlichny operator depends on the simultaneous coherence of a large number of these correlators. This makes it especially sensitive to accumulated imperfections across the subsystem. As a consequence, two processors with comparable gate fidelities can exhibit markedly different collective quantum performance, a difference that cannot be inferred from conventional benchmarking metrics alone. This is illustrated even within the same architecture: despite comparable single- and two-qubit gate fidelities, \texttt{AQT~IBEX~Q1} consistently sustains higher normalized Svetlichny violations than \texttt{AQT~Marmot} across most of the explored range (Fig.~\ref{fig:Normalized}). This gap is not captured by quantum volume, which is identical for both devices (128, Table \ref{tab:hardware}) despite their measurably different Svetlichny performance, reflecting that quantum volume, based on random circuit sampling, is not sensitive to this collective limitation. The mitigation analysis (Fig.~\ref{fig:MitigationHeatmap}) reinforces this: comparably modest gains on both processors indicate that readout errors make a comparably small contribution to the observed degradation on both processors, so the residual gap reflects device-specific differences that neither quantum volume nor gate-fidelity benchmarks are designed to detect. Collective performance therefore also depends on device-specific factors beyond average gate quality alone, a conclusion further supported by recent cross-platform benchmarking studies~\cite{Linke2017,Zhu2022} showing that system-level performance metrics can differ substantially from comparisons based solely on gate fidelities.

The readout mitigation analysis provides additional insight into the physical origin of performance degradation. For both AQT trapped-ion processors, \texttt{Marmot} and \texttt{IBEX~Q1}, local and correlated mitigation produce only modest corrections throughout the explored range, indicating that measurement imperfections are not the dominant limitation on either device. For the superconducting processor \texttt{IQM~EQE1}, the mitigation gain grows by nearly an order of magnitude between $N=2$ and $N=10$, and local mitigation recovers Svetlichny violations up to approximately $N=8$ qubits. This strong recovery indicates that a substantial fraction of the apparent collapse of multipartite nonlocality under raw execution conditions originates from measurement noise rather than from the destruction of the underlying quantum correlations themselves. This result is consistent with the known sensitivity of superconducting readout channels to crosstalk and residual correlations between measurement lines~\cite{Bravyi2021, Nation2021}, and highlights the importance of scalable mitigation strategies for reliable multipartite benchmarking on superconducting platforms.

Importantly, the absence of a Svetlichny violation observed in the demonstrations does not necessarily imply the total disappearance of multipartite quantum correlations: part of the nonlocal structure may remain physically present while becoming effectively hidden by readout noise. This distinction is consistent with the interpretation of these results as simulating genuine multipartite nonlocality, as discussed in Sec.~\ref{sec:GMNL}: given the inherent loopholes affecting co-located qubits, the observed loss of violation reflects a loss of the signal used to simulate this effect, rather than necessarily a loss of the underlying non-classical resource itself, which, following~\cite{Granda_2025} (Sec.~VI.B), can alternatively be captured through a contextuality-based reading of the same correlations. This distinction has direct practical implications for the interpretation of benchmarking results on NISQ devices, where measurement imperfections can systematically mask the actual collective quantum capability of the processor.

The comparison between local and correlated mitigation further reveals a fundamental scalability tension inherent to large-scale multipartite benchmarking. Correlated mitigation provides the most accurate correction by reconstructing the complete $2^N \times 2^N$ assignment matrix, but its calibration overhead grows exponentially with subsystem size, limiting its applicability to $N \leq 6$ in this work. The complete Svetlichny evaluation itself also requires $2^N$ measurement configurations. Thus, while the present ten-qubit implementation demonstrates the practical feasibility and discriminating power of the protocol at nontrivial subsystem sizes, further scaling will require sampling strategies or reduced-measurement formulations.

Local mitigation, by contrast, remains tractable at all scales and still recovers a substantial fraction of the nonclassical region in the superconducting architecture. Developing scalable mitigation strategies that interpolate between these two extremes (for instance, through structured approximations of the assignment matrix~\cite{Bravyi2021} or model-free approaches~\cite{VanDenBerg2022}) constitutes an important direction for future multipartite benchmarking protocols.

Finally, although the present work focuses on GHZ states and Svetlichny inequalities, the benchmarking protocol is not tied to this particular choice of resource state. We have previously studied how multipartite nonlocality is distributed and scales across other circuit families, including Clifford, Clifford+\texttt{T}, and Haar-random circuits~\cite{Granda_2025}. The same demonstration approach used here, evaluating Svetlichny-type correlators on real hardware, applies directly to those settings. Graph states and linear cluster states are a further natural extension: their Bell inequalities show improved noise resistance at large $N$~\cite{Bonsel2024}, and comparing them against GHZ-based protocols on the same hardware would clarify how entanglement topology and connectivity shape the robustness of multipartite nonlocal correlations in future large-scale quantum architectures. More broadly, the present results establish a general benchmarking strategy in which a common multipartite workload is specified independently of the target hardware and subsequently instantiated across different processors to probe their collective quantum performance on a common scale.

\begin{acknowledgments}
The authors gratefully acknowledge Felix Rohde and the entire AQT team for their fruitful discussions.

The authors gratefully acknowledge the use of the quantum systems \texttt{IQM~EQE1} and \texttt{AQT~Marmot}, operated by the Leibniz Supercomputing Centre (LRZ), for providing the computational resources for this work.
\end{acknowledgments}

\section*{Data Availability}
The data that support the findings of this article are not
publicly available. The data are available from the authors
upon reasonable request.

\appendix

\section{Quantum hardware platforms}
\label{sec:hardware}

The demonstrations were performed on three quantum processors spanning two fundamentally different architectures: the trapped-ion processors \texttt{Marmot} and \texttt{IBEX~Q1}, both developed by Alpine Quantum Technologies (AQT)~\cite{AQT_Marmot,AQT_IBEX}, and the superconducting processor \texttt{EQE1} developed by IQM~\cite{LRZ_EuroQExa}. \texttt{Marmot} and \texttt{EQE1} are operated through the Leibniz Supercomputing Centre (LRZ), while \texttt{IBEX~Q1} is accessed via AQT's own cloud service (ARNICA) from Innsbruck, Austria. Table~\ref{tab:hardware} summarizes the main hardware characteristics of the three processors.

\begin{table*}[]
\centering
\caption{Comparison of selected hardware characteristics of the three quantum processors used in this work. The reported characteristics correspond to the hardware configurations used during the demonstrations in August 2026.}
\label{tab:hardware}
\begin{tabular}{lccc}
\toprule
Property & \texttt{AQT Marmot} & \texttt{AQT IBEX Q1} & \texttt{IQM EQE1} \\
\midrule
Physical qubits
    & 16 (15)
    & 12
    & 54 \\

Connectivity / topology
    & All-to-all
    & All-to-all
    & Square lattice, tunable couplers \\

Single-qubit gate fidelity (\%)\tnote{a}
    & $99.9471\pm0.0067$
    & $99.9588\pm0.0045$
    &  $ \geq 99.93$ \\

Two-qubit gate fidelity (\%)\tnote{a}
    & $97.3395\pm1.7169$
    & $99.02\pm0.18$
    & median: $99.1441$ \\

$T_1$
    & $1.168\pm0.007$ s
    & $1.168\pm0.007$ s
    & median: $40.78\,\mu$s \\

$T_2$
    & $0.1632\pm0.0442$ s
    & $0.1632\pm0.0442$ s
    & median: $8.45\,\mu$s \\



Readout time
    & $1500\,\mu$s
    & $1500\,\mu$s
    & -- \\

Single-qubit gate duration
    & $45\,\mu$s
    & $45\,\mu$s
    & -- \\

Two-qubit gate duration
    & $335\,\mu$s
    & $335\,\mu$s
    & -- \\

Quantum volume
    & 128
    & 128
    & 64 \\

Host / access
    & LRZ
    & AQT (ARNICA)
    & LRZ \\

\bottomrule
\end{tabular}

\begin{tablenotes}
\small
\item[a] Fidelities are as reported by each provider and are not directly comparable across platforms. The AQT values are true averages from randomized benchmarking (1-qubit RB: 50 shots per circuit, 15 random circuits per Clifford length, lengths $\{1,100,200,400,600\}$; 2-qubit RB: 25 shots per circuit, 15 random circuits per Clifford length, lengths $\{1,16,32\}$, performed on each of the 12 qubit pairs), with the quoted uncertainty reflecting this statistics. The IQM value is a median, which is less sensitive to outliers than a mean and can differ substantially from it when the underlying fidelity distribution has a large spread, as is typical for superconducting platforms.
\end{tablenotes}
\end{table*}

Because the number of correlators required for the generalized Svetlichny operator grows exponentially with the subsystem size (specifically, $2^N$ correlators are needed, corresponding to 128, 256, 512, and 1024 correlators for $N=7,8,9,$ and $10$, respectively), the statistical requirements become increasingly demanding at larger qubit numbers. Because the \texttt{AQT Marmot} platform was limited to 200 shots per job, each demonstration repetition was constructed from ten independent 200-shot jobs, whose counts were aggregated to obtain an effective shot budget of 2,000 shots. For \texttt{AQT IBEX Q1} and \texttt{IQM EQE1}, the same effective budget of 2,000 shots was used; since the provider's interface permitted this, a single job of 2,000 shots was submitted directly. This 2,000-shot procedure was independently repeated five times for each device and subsystem size \(N\). The reported \(S_N\) is the mean of these five repetition-level estimates, and the 95\% confidence intervals are calculated from their variability using a Student’s \(t\)-distribution with four degrees of freedom.

\section{State preparation: GHZ circuits}
\label{sec:ghz}

To investigate the scalability of genuine multipartite nonlocality, we employ Greenberger--Horne--Zeilinger (GHZ) states as the reference multipartite entangled resource. GHZ states are particularly suitable for this purpose because they achieve the maximal quantum violations of generalized Svetlichny inequalities~\cite{Seevinck_Svetlichny_PRL} and therefore constitute a natural benchmark for assessing the preservation of high-order nonlocal correlations in quantum processors.

For an $N$-qubit system, the generalized GHZ state is

\begin{equation}
\label{eq:GHZ_state}
|\mathrm{GHZ}_{N}^{\pm}\rangle=
\frac{1}{\sqrt{2}}
\left(
|0\rangle^{\otimes N}
\pm
|1\rangle^{\otimes N}
\right).
\end{equation}

Following the original Svetlichny formalism, the local observables are restricted to the $x$-$y$ plane of the Bloch sphere and are defined as

\begin{equation}
A_k^{i_k}
=
\cos\alpha_k^{i_k}\,\sigma_x
+
\sin\alpha_k^{i_k}\,\sigma_y,
\end{equation}

\noindent where $k=1,\ldots,N$ labels the qubit, $i_k\in\{1,2\}$ identifies the measurement setting, and $\alpha_k^{i_k}$ denotes the corresponding measurement angle. For the GHZ state, the multipartite correlation function takes the form

\begin{equation}
E(i_1,\ldots,i_N)
=
\pm
\cos
\left(
\alpha_1^{i_1}
+\cdots+
\alpha_N^{i_N}
\right),
\end{equation}

\noindent where the sign coincides with that chosen in Eq.~\ref{eq:GHZ_state}. As shown in Ref.~\cite{Seevinck_Svetlichny_PRL}, maximal violations of the generalized Svetlichny inequalities are obtained by choosing the measurement settings

\begin{equation}
\left(
\alpha_1^{1},
\alpha_2^{1},
\ldots,
\alpha_N^{1}
\right)
=
\left(
\frac{\pi}{4},
0,
\ldots,
0
\right),
\end{equation}

\noindent and

\begin{equation}
\left(
\alpha_1^{2},
\alpha_2^{2},
\ldots,
\alpha_N^{2}
\right)
=
\left(
\frac{3\pi}{4},
\frac{\pi}{2},
\ldots,
\frac{\pi}{2}
\right).
\end{equation}

The asymmetric choice of measurement angles, in which the first qubit plays a distinguished role, follows directly from the analytical construction of the generalized Svetlichny inequalities and yields the maximal theoretical violation for GHZ-states~\cite{Seevinck_Svetlichny_PRL}. Throughout this work, we employ the positive-sign convention corresponding to the original theoretical prescription.

Importantly, no numerical optimization of the measurement settings was performed. All demonstrations were executed using prescribed by the Svetlichny formalism, ensuring a scalable, reproducible, and architecture-independent benchmarking protocol.

\begin{figure*}[]
\centering
\includegraphics[width=0.9\linewidth]{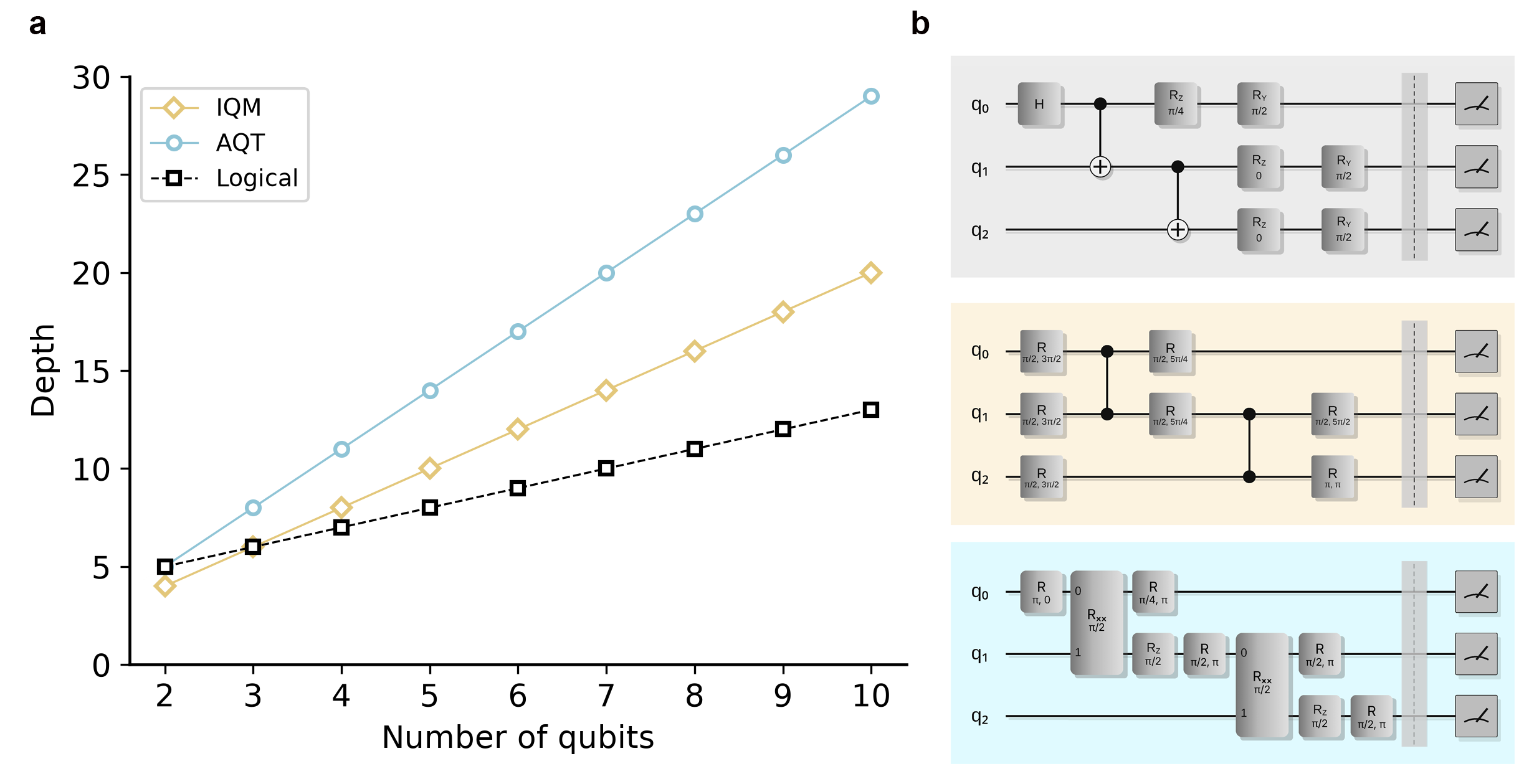}
\caption{GHZ-state preparation circuit: (a) Logical circuit (black), IQM (cream), and AQT (blue) transpiled implementations. (b) $N=3$ example: logical circuit (top, black), IQM implementation (middle, cream), AQT implementation (bottom, blue).}
\label{fig:GHZ}
\end{figure*}

For the demonstrations, the logical GHZ circuits were constructed using standard universal quantum gates: a Hadamard gate (\texttt{H}) acting on the first qubit followed by a cascade of controlled-NOT (\texttt{CNOT}) gates distributing entanglement across the register. The resulting logical circuits were subsequently transpiled independently for each quantum processor according to the corresponding native gate sets and hardware constraints. The resulting compilation overhead, across all subsystem sizes considered in this work, is quantified in  Table~\ref{tab:circuit_complexity} and visualized in Fig.~\ref{fig:GHZ}(a), showing how circuit depth scales with $N$ for the logical circuit and its transpiled implementations on the IQM and AQT native gate sets; panel~\ref{fig:GHZ}(b) shows one concrete instance of this comparison at $N=3$, with the logical circuit and both transpiled implementations drawn out explicitly.

This strategy was intentionally adopted to guarantee identical logical input circuits across all devices. In this way, the observed differences in performance reflect not only the physical hardware itself, but also the combined effects of connectivity constraints, compilation overhead, and transpilation efficiency.

\section{Readout error mitigation}
\label{sec:mitigation}

In current NISQ processors, imperfections in the measurement process can cause the recorded bitstring to differ from the actual computational basis state, directly affecting the estimated probability distributions from which multipartite correlation functions are evaluated. Since multipartite Bell-type inequalities are evaluated from probability distributions estimated in the demonstrations, readout errors directly affect the measured correlation functions and can significantly reduce the observed nonlocality.

The effect of measurement noise can be described as a classical stochastic process acting on the ideal probability distribution. Let $\mathbf{p}_{\mathrm{true}}$ denote the probability vector associated with the ideal measurement outcomes and $\mathbf{p}_{\mathrm{obs}}$ the observed distribution. The two distributions are related through an assignment matrix $A$ according to $\mathbf{p}_{\mathrm{obs}} = A \, \mathbf{p}_{\mathrm{true}}$, where each matrix element $A_{yx}=P(y|x)$ represents the probability of observing outcome $y$ when the actual state is $x$. Readout error mitigation aims to estimate the inverse action of this process and recover a more accurate approximation of the underlying probability distribution~\cite{Bravyi2021}.

In this work, two measurement error mitigation strategies available in the \texttt{Qiskit Experiments} framework~\cite{Qiskit, Nation2021} were investigated: \emph{Local Readout Error Mitigation} and \emph{Correlated Readout Error Mitigation}. These approaches differ in the assumptions made regarding the structure of the measurement noise and consequently in their calibration overhead and scalability. For each subsystem size, the calibration circuits were executed within the same demonstration session as the nonlocality circuits to reduce the impact of temporal calibration drift.

\paragraph{Local readout error mitigation.}
The local approach assumes that measurement errors occur independently on each qubit and that no correlations exist between different readout channels. Under this approximation, the global assignment matrix factorizes as $A = A_1 \otimes A_2 \otimes \cdots \otimes A_N$, where each single-qubit assignment matrix is given by

\begin{equation}
A_i =
\begin{pmatrix}
P(0|0) & P(0|1) \\
P(1|0) & P(1|1)
\end{pmatrix}.
\end{equation}

The calibration procedure requires only two calibration circuits, corresponding to the preparation of the states $|0\rangle^{\otimes N}$ and $|1\rangle^{\otimes N}$. Under the locality assumption, the resulting measurements are used to estimate the individual assignment probabilities for each qubit. Once the local calibration matrices are constructed, the mitigated probability distribution is obtained by solving

\begin{equation}
\mathbf{p}_{\mathrm{mit}}
=
A^{-1}
\mathbf{p}_{\mathrm{obs}},
\end{equation}

\noindent where the inversion is performed numerically in a least-squares or pseudoinverse sense in order to obtain a physically consistent probability distribution~\cite{Bravyi2021, Nation2021}. A major advantage of this method is its excellent scalability. While a complete characterization of the readout process would require a number of calibration parameters growing exponentially with the number of qubits, the local approximation reduces the calibration cost to a quantity that scales linearly with system size.

\paragraph{Correlated readout error mitigation.}
The correlated approach removes the independence assumption and explicitly models joint measurement errors involving multiple qubits. In this case, the complete assignment matrix $A\in\mathbb{R}^{2^N\times 2^N}$ is reconstructed from all conditional probabilities

\begin{equation}
A_{yx}=P(y|x),
\end{equation}

\noindent where, as above, $y$ and $x$ range over all computational basis states of the $N$-qubit register, corresponding to the observed and true states, respectively. This allows the mitigation procedure to capture measurement crosstalk and residual correlations between readout channels.

To estimate all matrix elements, calibration circuits must be executed for every computational basis state $|00\cdots0\rangle, |00\cdots1\rangle, \dots, |11\cdots1\rangle$, resulting in a total of $2^N$ calibration circuits. Although this strategy generally provides a more accurate characterization of the measurement process, its calibration overhead grows exponentially with the number of qubits, making it impractical for large-scale demonstrations.

For this reason, correlated mitigation was restricted to systems of up to six qubits for all three processors, while local mitigation was employed up to ten qubits for all three processors: \texttt{Marmot}, \texttt{AQT~IBEX~Q1}, and \texttt{EQE1}. This triple comparison allows us to assess the impact of readout errors on the observed Svetlichny violations and to evaluate the extent to which measurement imperfections contribute to the degradation of multipartite nonlocal correlations across all three processors.

To quantify the effect of mitigation, we define the relative improvement

\begin{equation}
\label{e:delta_mit}
\Delta_{\mathrm{mit}} = \frac{\mathcal{V}_N^{\mathrm{mit}} - \mathcal{V}_N^{\mathrm{raw}}}{\mathcal{V}_N^{\mathrm{raw}}} \times 100\%,
\end{equation}

\noindent where $\mathcal{V}_N^{\mathrm{raw}}$ and $\mathcal{V}_N^{\mathrm{mit}}$ denote the normalized Svetlichny violation under raw and mitigated conditions, respectively, for either the local or correlated strategy.

\bibliography{Bibliography_base}

\end{document}